\documentclass[draft]{agujournal}
\draftfalse

\journalname{<JGR-Space Physics>}

\begin{document} 

\title{Magnetic Shear Damped Polar Convective Fluid Instabilities}

\authors{Jyoti K Atul\affil{1}, Rameswar Singh\affil{1}, Sanjib Sarkar\affil{2}, Oleg V Kravchenko\affil{3,4,5}, Sushil K Singh\affil{6}, Prabal K Chattopadhyaya\affil{1}, Predhiman K Kaw\affil{1}}

\affiliation{1}{Institute for Plasma Research, Bhat, Gandhinagar, India.}

\affiliation{2}{Institute of Plasma Physics, Chinese Academy of Sciences, Anhui, People's Republic of China.}

\affiliation{3}{Scientific and Technological Center of Unique Instrumentation,Russian Academy of Sciences,Moscow,Russian Federation.}

\affiliation{4}{Kotel'nikov Institute of Radioengineering and Electronics, Russian Academy of Sciences, Moscow, Russian Federation.}

\affiliation{5}{Department of Higher Mathematics, Bauman Moscow State Technical University, Moscow, Russian Federation.}

\affiliation{6}{Department of Physics, Magadh University, Bodh Gaya, India.}  


\correspondingauthor{Jyoti K Atul}{jkatulphysics@gmail.com}

\begin{keypoints}
\item Convective Fluid Instabilities 
\item Auroral Plasma 
\item Blob Dynamics 
\end{keypoints}

\justify

\begin{abstract}
The influence of the magnetic field shear is studied on the $E\times B$ (and/or Gravitational) and the Current Convective Instabilities (CCI) occuring in the High latitude F-layer ionosphere. It is shown that magnetic shear reduces the growth rate of these instabilties.
The magnetic shear induced stabilization is more effective at the larger scale sizes ($\geq$ tens of kilometers) while at the scintillation causing intermediate scale sizes ($\sim$ a few kms), the growth rate is largely unaffected. The eigen mode structure is localised about a rational surface due to finite magnetic shear and has broken reflectional symmetry due to centroid shift of the mode  by equilibrium parallel flow or current. 
\end{abstract}

\section{Introduction}

High latitude ionospheric plasma dynamics
reveals many interesting phenomenon. It includes the production and
convection of large-scale plasma enhancements such as \char`\"{}patches\char`\"{} and \char`\"{}blobs,\char`\"{}, the acceleration and heating of ionospheric ions into the magnetosphere and also vivid auroral displays which are considered to be the manifestations of the substorm dynamics. The generation and convection mechanism  of these patches and blobs are quite interesting among
the near-earth space events. These large-scale (macro-scale)
plasma enhancements are of global origin and have been characterised
as patches (in the polar cap) and blobs (at the auroral latitudes). Intense observational and simulation studies shows that these patches drift to long distances and long periods of time while retaining their distinct identity. In addition to it, mesocale irregularities are widely observed in association with these patches throughout the polar cap region \citet{kelley2009earth}.

Similar blob morphology and propagating coherent structures with different scales, have also been observed in solar photosphere \citet{fundamenski2007relationship} and in the edge/scrape-off layer of toroidal plasma fusion devices such as Tokamaks \citet{zweben2002edge}, \citet{bisai2005formation}, \citet{krasheninnikov2008recent},
\citet{xu2009blob}. In Tokamaks, these high density blobs are propelled through the background plasma by a charge polarization induced by magnetic curvature, gradient drifts and a corresponding EXB radial convection. The Tokamak blob dynamics is believed to dominate the scrape-off layer transport, possibly leading to impurity generation and serious wall erosion.  

For the High Latitude bolb scenario, the density of these ionospheric structures ranges from two to ten times the background density therby producing deleterious effects on communications systems through scintillation of RF waves. These electron density irregularities may disrupt Very High Frequency (VHF), Ultra High Frequency (UHF), and Global Navigation Satellite Systems (GNSS) at L-band frequencies. Infact, radio signals gets disturbed or interrupted while their propagation through these ionospheric irregularities, which are often associated with large density gradients resulting in amplitude and phase fluctuations. Thus, these disturbances leads to degraded performance for GNSS receivers and occasional loss of navigation solutions \citet{huba1988simulations}, \citet{mitchell2005gps},   \citet{moen2013space}, \citet{wang2016comparison}.

Ionospheric tomography has been upgraded from 2-D simulations to Advanced Global 4-D ionospheric imaging in last decades. Ionospheric imaging techniques involves the usage of  integrated  electron density measurements (Total Electron Content/TEC) to develop 2-D, 3-D and 4-D electron density maps \citet{bust2008history}. Nowadays, Ionospheric imaging techniques are being used to probe the dominant scintillation zones across the equatorial as well as high latitude regions, in order to develop efficient scintillation forecasting models \citet{wernik2003ionospheric}, \citet{ledvina2004temporal},
\citet{burston2012imaging}, \citet{priyadarshi2015review}. It turns out that electron density profile measurements are crucial to access the  horizontal and vertical distribution of the global plasma structure and its temporal evolution.

Multi-instrumental co-ordinated observations of these electron density structures have been made through ground based EISCAT incoherent radars, SuperDARN HF coherent radars, in-situ rockets, optical all sky imagers and GPS scintillation measurements \citet{oksavik2006observations}, \citet{yin2009imaging},
\citet{oksavik2010entry}, \citet{oksavik2012situ},
\citet{zhang2013direct}, \citet{hosokawa2013two},
\citet{jin2014gps}, \citet{spicher2015observation}, 
\citet{jin2016statistical}, \citet{clausen2016gps},
\citet{lamarche2017radar}. Among the polar and auroral sectors of the globe, a broad network of coherent and incoherent radars provide continuous monitoring of the high-latitude ionospheric plasma convection patterns, structurisation and reorganisation processes of the plasma patches. High latitude Incoherent scatter radar (ISR) facilities include Sondrestrom in Greenland, EISCAT in northern Scandinavia, EISCAT Svalbard Radar (ESR) on Svalbard, Irkutsk in Russian Federation, Advanced Modular Incoherent Scatter Radar at Fairbanks, Alaska, Resolute Bay ISR (RISR-N) in northern Canada \citet{dahlgren2012space}, RISR-C \citet{gillies2016first} whereas HF backscatter radar facilities includes Super Dual Auroral Radar network (SuperDARN) Hankasalmi radar in Finland and the SuperDarn Kodiak radar in Alaska. 

There has also been extensive theoretical and computational studies on high-latitude structure and turbulence. Several primary plasma instabilities have been proposed as the cause of scintillation inducing irregularites associated with plasma patches. The bulk of these studies has focused on Gradient Drift Instability (GDI)
\citet{sojka1998gradient}, \citet{guzdar1998three},    \citet{gondarenko1999gradient}, 
\citet{gondarenko2001three},
\citet{gondarenko2003structuring}, \citet{gondarenko2003structure},
\citet{gondarenko2004density}, \citet{gondarenko2004plasma}, \citet{gondarenko2006nonlinear}, \citet{gondarenko2006simulations}, 
Current Convective instabilty (CCI) 
\citet{ossakow1979current}, \citet{chaturvedi1979nonlinear},
\citet{huba1980influence}, \citet{chaturvedi1981current}, \citet{huba1984long}, \citet{huba1986effect}, \citet{chaturvedi1994effects},
Kelvin-Helmholtz instabilty (KHI)
\citet{keskinen1988nonlinear} 
\citet{gondarenko2006nonlinear},
\citet{carlson2007case}, \citet{oksavik2010entry}.
and turbulent processes in the high latitude F-layer ionosphere
\citet{burston2009correlation}, \citet{burston2010turbulent}.

Theoretically, it is beleived that 3D analytical treatments in the collisional regime for the GDI studies shows that dynamics parallel to the magnetic field are stabilizing at long wavelengths \citet{chaturvedi1987interchange}. Based upon this theoretical motivation, numerical simulations have verified the structuring processes in the plasma patches. \citet{guzdar1998three} shows that inclusion of 3D effects, nonlinear evolution is dominated by the generation of mesoscales and the deletarious long wavelengths are supressed.Later, \citet{gondarenko1999gradient} extended the investigation to include the combined effect of parallel dynamics and the inertial effects. The nonlinear simulation shown that the initial cross-field elongated structures were unstable to secondary KHI further leading to breakdown of structures into sub-structures. Thus, these interplays lead to a complex nonlinear state consisting of density and potential fluctuations packed in multiple shear layers in the system.

Through a more sophisticated modelling of the plasma patch, \citet{gondarenko2001three}, \citet{gondarenko2003structure} have shown that these density irregularities doesn't remain localized in the edges but progressively penetrate in the entire plasma patch during the nonlinear evolution process. Further, ion neutral collisions play a major role in the determination of both saturation levels for the density and potential fluctuations and the nature of turbulent spectra. Thus, it turns out that the combined effects of the parallel dynamics, nonlinear ion inertial effects with the altitude dependent ion neutral collision frequency unify the natural GDI and KHI sources leading to small scale sub-structures generation in the polar cap plasma patches. However, in these earlier studies, the drive was assumed to be composed of $E\times B$ and neutral wind velocity and constant in time. The study was further allowed to include the variable drive obtained from ionospheric module of the NRL global MHD code simulation of a real substorm \citet{fedder1995topological}, \citet{sojka1997driving}.     

The magnitudes and spectral characteristics of the density and electric field fluctuations arising due to primary and secondary instabilities were investigated in detail in \citet{gondarenko2004density}, \citet{gondarenko2004plasma}. It was concluded that a multistep process involving a primary GDI, secondary KHI and tertiary shear flow instabilities, are responsible for the nonlinear structurisation process. These nonlinear processes further leads to the existence of mesoscale structures on the edges as well as interiors of the patches.   
The investigation was further repeated using the primary GDI and primary KHI \citet{gondarenko2006nonlinear}, \citet{gondarenko2006simulations} and it was found that the shear layer generation by the primary KHI is more stronger than that due to secondary KHI. Moreover, the basic structuring process was categorized into four groups. Further, it was concluded that scintillation index/normalised irregularity index values are found to be weaker than those with no shear. It turns out patch structurisation by the primary GDI and high plasma density in the patch can cause intense scintillation.   

Recently, \citet{burston2016polar} investigated the role of four primary plasma instabilties in the generation of phase scintillation associated with the polar cap plasma patches using Dynamic Explorer 2 Satellite data. Further, GDI, CCI, KHI and small scale turbulence processes were examined statistically. These studies suggested inertial turbulence instabilty to be the dominant process, followed by the inertial gradient drift, collisional tubulence and the collisional shortwave CCI. However, the other processes, such as KHI, collisional GDI and inertial shortwave CCI, were found to be relatively unimportant to give rise to GPS scintillation.

With our prime motivation in investigating the role of such convective fluid  plasma multiple instabilities in the generation of scintillation inducing irregularities, local and global analysis has been carried out for the $E\times B$ (and/or Gravitational) and CCI primary instabilities under the influence of sheared magnetic field in the slab geometry. 
The plan of the paper is as follows. A general description of the unstable convective plasma motions, under the local approximation, in presence of a transverse inhomogeneity; with contributions to the growing fields coming from the combined effect of the presence (in equilibrium) of a parallel current, acceleration due to gravity and a transverse electric field, is presented in section 2. Further, in section 3, the treatment is extended to include magnetic field shear in order to address the nonlocal problem for these convective fluid instabilities. Lastly, the conclusion is delivered in section 4.

\section{Local analysis}

\begin{figure}[ht!]
\centerline{\includegraphics[scale=0.8]{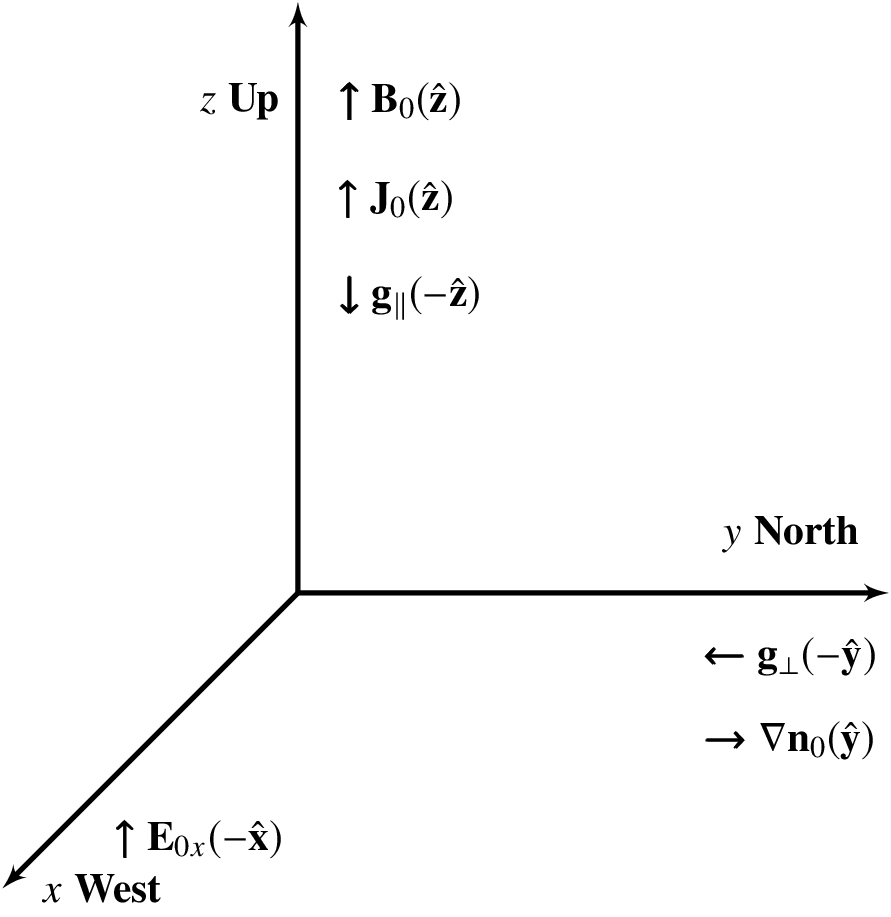}}

\caption{(Colour online) Field geometry for the High latitude F-layer Ionosphere.}
\label{fieldmat14}
\end{figure}

The situation encountered at the F-region altitudes of
high latitude ionosphere under the diffuse auroral conditions,is considered. The assumptions are as described below. In the co-ordinate system considered, Z-axis is aligned with the magnetic field of the earth $B\left(\hat{\boldsymbol{z}}\right)$
and Y-axis points in the northward direction whereas X-axis points
in the westward direction. In equilibrium, an electric field $E_{0x}\left(-\hat{\boldsymbol{x}}\right)$,
a density gradient $\frac{d}{dy}n_{0}\hat{\boldsymbol{y}}$ and a
current $J_{0}\left(\hat{\boldsymbol{z}}\right)$ are assumed to exist.
The acceleration due to earth's gravity also has a component transverse
to the magnetic field $g_{\perp}\left(-\hat{\boldsymbol{y}}\right)$.
A relative equilibrium drift of ions $V_{0}\left(\hat{\boldsymbol{z}}\right)$ over electrons is assumed to simulate the zero-order field-aligned currents in a frame of reference in which the electrons are at rest. The temperature effects are neglected $\left(T_{e}=T_{i}=0\right)$ for both the electron and ion species. Thus, the treatment is valid at the transverse wavelengths longer than the diffusion cutoff (~few hundred meters). Electron inertia is also ignored. Further, the vector quantities are written in bold, ion properties are indicated with capital letters and electron properties are indicated in small letters. Equilibrium quantities are suffixed with $0$ and perturbations are suffixed with $1$. The detailed field geometry under consideration is given in Fig.\,\ref{fieldmat14}. 

The basic equations for ion and electron dynamics consist of respective
continuity and momentum equations as given below. 
\begin{equation}
\frac{\partial}{\partial t}N+\boldsymbol{\nabla}.\left(N\boldsymbol{V}\right)=0\label{eq:ni}
\end{equation}

\begin{equation}
\left(\frac{\partial}{\partial t}+\boldsymbol{V}.\boldsymbol{\nabla}+\nu_{in}\right)\boldsymbol{V}=\frac{e}{M}\boldsymbol{E}+\left(\boldsymbol{V}\times\boldsymbol{\Omega}_{i}\right)+\boldsymbol{g}
\end{equation}

\begin{equation}
\frac{\partial}{\partial t}n+\boldsymbol{\nabla}.\left(n\boldsymbol{v}\right)=0
\end{equation}

\begin{equation}
\left(\frac{\partial}{\partial t}+\boldsymbol{v}.\boldsymbol{\nabla}+\nu_{ei}\right)\textbf{v}=-\frac{e}{m}\boldsymbol{E}+\left(\boldsymbol{v}\times\boldsymbol{\Omega}_{e}\right)
\end{equation}
In the low collisionality regime $\nu_{in}\ll\Omega_{i}$ the zero-order
ion drift velocity is given as

\begin{equation}
\boldsymbol{V}_{0}=\frac{\nu_{in}}{\Omega_{i}}\frac{c}{B_{0}}\boldsymbol{E}_{0\perp}+\frac{\nu_{in}}{\Omega_{i}^{2}}\boldsymbol{g}_{\perp}+\frac{1}{\Omega_{i}}\boldsymbol{g}_{\perp}\times\hat{\boldsymbol{z}}+\frac{j_{0\parallel}}{n_{o}e}\hat{\boldsymbol{z}}
\end{equation}
which is made up of sum of electric Pedersen drift or collision modified EXB drift (first term), gravitational Pedersen drift or collision modified gravitational drift (second term), pure gravitational drift (third term) and parallel mean flow (fourth term). Here zero-order
ion drift velocity is transformed to a reference frame drifting with the $E\times B$ drift speed $V_{E}=\left(\frac{c}{B_{0}}\boldsymbol{E}_{0\perp}\times\hat{\boldsymbol{z}}\right)$.
Assuming that the perturbations vary as $f=f_{0}e^{i\left(\boldsymbol{k}.\boldsymbol{r}-\omega t\right)}$
and that the perturbed electric fields are electrostatic $\boldsymbol{E}_{1}=-\boldsymbol{\nabla}\phi_{1}$
where $\omega$ is a characteristic frequency and further assuming
that $\left(\omega-k\cdot V_{0}\right)\sim\nu_{in}\ll\Omega_{i}$
the perturbed perpendicular ion velocitiy is given as

\begin{equation}
\boldsymbol{V}_{1\perp}=-i\frac{c}{B_{0}}\left[\frac{\lbrace\nu_{in}-i\left(\omega-\boldsymbol{k}.\boldsymbol{V}_{0}\right)\rbrace}{\Omega_{i}}\boldsymbol{k}_{\perp}+\left(\boldsymbol{k}_{\perp}\times\hat{\boldsymbol{z}}\right)\right]\phi_{1}\label{eq:vperpi}
\end{equation}
which is made up of sum of Pedersen drift (first term), inertial drift
(second) and $E\times B$ drift (third term). The perturbed parallel
ion velocity becomes

\begin{eqnarray}
\boldsymbol{V}_{1z} & = & -i\frac{c}{B_{0}}\left[\frac{\Omega_{i}}{\lbrace\nu_{in}-i\left(\omega-\boldsymbol{k}.\boldsymbol{V}_{0}\right)\rbrace}\boldsymbol{k}_{z}\right]\phi_{1}\label{eq:vpari}
\end{eqnarray}
Similarly, for electrons, the zero-order electron drift velocity is
given as
\begin{equation}
\boldsymbol{v_{0}}=0
\end{equation}
The perturbed electron velocity is composed up of sum of Pedersen drift, $E\times B$ drift and parallel velocity perturbation as follows.

\begin{equation}
\boldsymbol{v_{1}}=i\frac{c}{B_{0}}\left[\frac{\nu_{ei}}{\Omega_{e}}\boldsymbol{k}_{\perp}-\left(\boldsymbol{k}_{\perp}\times\hat{\boldsymbol{z}}\right)+\frac{\Omega_{e}}{\nu_{ei}}\boldsymbol{k}_{z}\right]\phi_{1}\label{eq:ve1}
\end{equation}
Substituting equations \ref{eq:vperpi} and \ref{eq:vpari} in the ion
continuity equation \ref{eq:ni} yields the following ion density
perturbation as

\begin{multline}
\frac{N_{1}}{N_{0}}=\frac{1}{\left(\omega-\boldsymbol{k}.\boldsymbol{V}_{0}\right)}\frac{c}{B_{0}}\\
\biggl[-\left\lbrace \frac{\lbrace\nu_{in}-i\left(\omega-\boldsymbol{k}.\boldsymbol{V}_{0}\right)\rbrace}{\Omega_{i}}k_{\perp}^{2}+\frac{\Omega_{i}}{\lbrace\nu_{in}-i\left(\omega-\boldsymbol{k}.\boldsymbol{V}_{0}\right)\rbrace}k_{z}^{2}\right\rbrace \\
-\left\lbrace \frac{\lbrace\nu_{in}-i\left(\omega-\boldsymbol{k}.\boldsymbol{V}_{0}\right)\rbrace}{\Omega_{i}}\boldsymbol{k}_{\perp}+\left(\boldsymbol{k}_{\perp}\times\hat{\boldsymbol{z}}\right)\right\rbrace .\boldsymbol{\epsilon}_{n}\biggr]\phi_{1}
\end{multline}
where the first term represents the density perturbation due to divergence of Pedersen plus inertial drift, the second term represents the effect of parallel compression and the third term the contribution to density perturbation due to convection of equilibrium density by the net pependicular drift. Similarly the electrons density perturbation is obtained as 

\begin{equation}
\frac{n_{1}}{n_{0}}=\frac{1}{\omega}\frac{c}{B_{0}}\left[i\left\lbrace \frac{\nu_{ei}}{\Omega_{e}}k_{\perp}^{2}+\frac{\Omega_{e}}{\nu_{ei}}k_{z}^{2}\right\rbrace +\left\lbrace \frac{\nu_{ei}}{\Omega_{e}}\boldsymbol{k}_{\perp}-\left(\boldsymbol{k}_{\perp}\times\hat{\boldsymbol{z}}\right)\right\rbrace .\boldsymbol{\epsilon}_{n}\right]\phi_{1}
\end{equation}
which can also be interpreted as for ion density perturbation except
that now the electron inertial drift is ignored. Finally, the
quasi-neutrality condition is matched to get the following local dispersion relation for convective mix-mode fluid instabilities

\begin{multline}
\left(\omega-\boldsymbol{k}.\boldsymbol{V}_{0}\right)\left[i\left\lbrace \left(\frac{\nu_{ei}}{\Omega_{e}}\right)^{2}k_{\perp}^{2}+k_{z}^{2}\right\rbrace +\frac{\nu_{ei}}{\Omega_{e}}\left\lbrace \frac{\nu_{ei}}{\Omega_{e}}\boldsymbol{k}_{\perp}-\left(\boldsymbol{k}_{\perp}\times\hat{\boldsymbol{z}}\right)\right\rbrace .\boldsymbol{\epsilon}_{n}\right]\phi_{1}\\
=\omega\biggl[\left\lbrace -\frac{\nu_{ei}}{\Omega_{e}}\frac{\lbrace\nu_{in}-i\left(\omega-\boldsymbol{k}.\boldsymbol{V}_{0}\right)\rbrace}{\Omega_{i}}k_{\perp}^{2}+\frac{\nu_{ei}}{\Omega_{e}}\frac{\Omega_{i}}{\lbrace\nu_{in}-i\left(\omega-\boldsymbol{k}.\boldsymbol{V}_{0}\right)\rbrace}k_{z}^{2}\right\rbrace \\
-\frac{\nu_{ei}}{\Omega_{e}}\left\lbrace \frac{\lbrace\nu_{in}-i\left(\omega-\boldsymbol{k}.\boldsymbol{V}_{0}\right)\rbrace}{\Omega_{i}}\boldsymbol{k}_{\perp}+\left(\boldsymbol{k}_{\perp}\times\hat{\boldsymbol{z}}\right)\right\rbrace .\boldsymbol{\epsilon}_{n}\biggr]\phi_{1}\label{dispersionlocal}
\end{multline}

Now from the local mix-mode dispersion relation, the familiar local growth rates for the CCI \citet{ossakow1979current} and $E\times B$ (and/or Gravitational) instabilities could be easily recovered. So, we ignore the fluctuating electron Pedersen drift (term proportional to $\left(\frac{\nu_{ei}}{\Omega_{e}}\right)^{2}$ in equation \ref{dispersionlocal}), approximate ion inertia term $\nu_{in}-i\left(\omega-\boldsymbol{k}.\boldsymbol{V}_{0}\right)\sim\nu_{in}$, further use the following assumptions $\boldsymbol{k}_{y}=0$, $\boldsymbol{V}_{0\perp y}=0$ along with the vectorial operations namely $\left(\boldsymbol{k}_{x}\times\hat{\boldsymbol{z}}.\boldsymbol{\epsilon}_{n}\right)=-{k}_{x}{\epsilon}_{n}$ and $\boldsymbol{k}_{x}.\boldsymbol{\epsilon}_{n}=0$   

Following \citet{huba1980influence} and substituting $\omega=\omega_{r}+i\gamma_{L}^{0}$,
the above equation leads to the local mix-mode real frequency given as

\begin{equation}
\omega_{r} = \frac{\frac{k_z^{2}}
{k_x}\frac{\Omega_{e}}{\nu_{ei}}\left[\left(\frac{\nu_{in}}{\Omega_{i}}\right)\left({\frac{cE_{0x}}{B_{0}}+\frac{g_{\perp y}}{\nu_{in}}}\right)+\frac{j_{0\parallel}}{n_{0}e}\frac{k_z}{k_x}\right]}
{\left(\frac{\Omega_{e}}{\nu_{ei}}+\frac{\Omega_{i}}{\nu_{in}}\right)\frac{k_z^{2}}{k_x^{2}}+\frac{\nu_{in}}{\Omega_{i}}}
\end{equation}

and yields the local mix-mode growth rate to be

\begin{equation}\label{mixmodeloc}
\gamma_{L}^{0} = \frac{\epsilon_{n}\left[\left(\frac{\nu_{in}}{\Omega_{i}}\right)\left({\frac{cE_{0x}}{B_{0}}+\frac{g_{\perp y}}{\nu_{in}}}\right)+\frac{j_{0\parallel}}{n_{0}e}\frac{k_z}{k_x}\right]}
{\left(\frac{\Omega_{e}}{\nu_{ei}}+\frac{\Omega_{i}}{\nu_{in}}\right)\frac{k_z^{2}}{k_x^{2}}+\frac{\nu_{in}}{\Omega_{i}}}
\end{equation}

Now by putting $V_{0z}=\frac{j_{0\parallel}}{n_{0}e}=0$ in equation \ref{mixmodeloc}, the familiar local $E\times B$ (and/or collisional Gravitational) growth rates is obtained as

\begin{equation}
\gamma_{L-EXB}^{0}=\epsilon_{n}\left[\frac{cE_{0x}}{B_{0}}+\frac{g_{\perp y}}{\nu_{in}}\right]\label{shearfreelocalexb}
\end{equation}

Similarly, by putting $V_{o\perp x}=\left(\frac{\nu_{in}}{\Omega_{i}}\right)\left(\frac{cE_{0x}}{B_{0}}+\frac{g_{\perp y}}{\nu_{in}}\right)=0$  in equation \ref{mixmodeloc},
the maximum growth rate for the local CCI which maximises for 

\begin{equation}
\Theta_{max}=\frac{k_z}{k_x}=\left[\frac{\nu_{in}}{\Omega_{i}}\left(\frac{\Omega_{e}}{\nu_{ei}}+\frac{\Omega_{i}}{\nu_{in}}\right)^{-1}\right]^{\frac{1}{2}}
\end{equation}
turns out to be

\begin{equation}
\gamma_{L-CCI}^{0}=\frac{\left(\frac{\epsilon_{n}j_{0\parallel}}{n_{0}e}\right)}{2\left(1+\frac{\Omega_{e}}{\Omega_{i}}\frac{\nu_{in}}{\nu_{ei}}\right)^{\frac{1}{2}}}\label{shearfreelocalcci}
\end{equation}

\section{Non Local Analysis}
Now equation \ref{dispersionlocal} for local dispersion relation representing convective mix-mode fluid instabilities  could also be written in the following form with a dispersion function

\begin{equation}
D\left[k_{x},k_{y},k_{z},\omega\right]\phi_{1}=0
\end{equation}
It is to be noted that, in the above subsection, the term  $\boldsymbol{k}_{\perp}.\boldsymbol{\epsilon}_{n}$ is retained
that is usually neglected in the local analysis under the assumption of $\boldsymbol{\epsilon}_{n}\ll\boldsymbol{k}_{\perp}$. In the nonlocal analysis in this subsection, however, this condition will be relaxed and further long wavelengths will be considered such that $\boldsymbol{\epsilon}_{n}\geq\boldsymbol{k}_{\perp}$.

\begin{figure}[ht!]
\centerline{\includegraphics[scale=0.7]{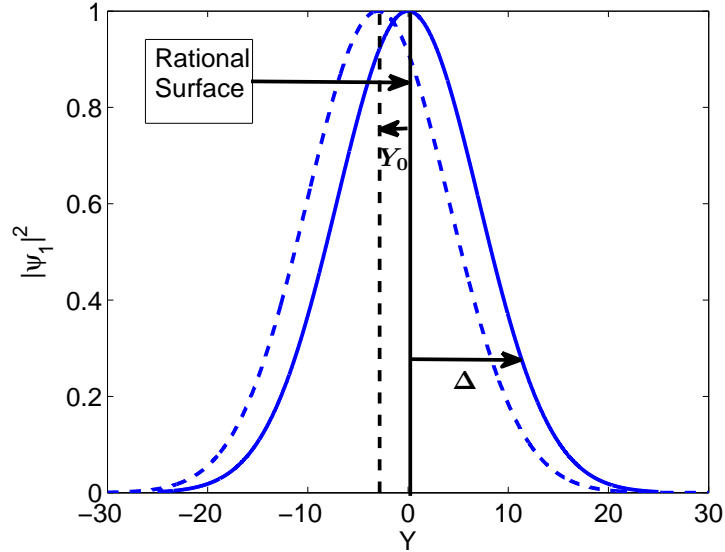}}

\caption{(Colour online) A generalised mode-cartoon for the representation of the mode which gets shifted off the rational
surface due to equilibrium parallel flow. Here $\Delta$ refers to mode width whereas $Y_{0}$ denotes mode shift}
\label{modecartoonfinal}
\end{figure}

The presence of a zero-order current $J_{0}\hat{\boldsymbol{z}}=n_{0}eV_{0}\hat{\boldsymbol{z}}$
in the system introduces a shear in the ambient magnetic field. Now
the field lines, though straight, are no longer parallel to each other,

\begin{equation}
\boldsymbol{B_{0}}=B_{0}\hat{\boldsymbol{z}}+B_{0x}(y)\hat{\boldsymbol{x}};B_{0}\gg B_{0x}
\end{equation}
Thus the perturbed quantities are also a function of $y$ now and
Fourier analysis in the Y-direction is not a valid procedure. So, one
can use Mikhilovskii prescription
\citet{chaturvedi225artificial}, \citet{chaturvedi1990eigenmode},\citet{mikhailovskii2013theory}, such that
\begin{equation}
k_{z}=k_{x}\frac{y}{l_{s}};k_{y}=-i\frac{\partial}{\partial y};k_{y}^{2}=-\frac{\partial^{2}}{\partial y^{2}}
\end{equation}
Further, Taylor expansion of $D$ in $k_{y}$ about $k_{y}=0$ is estimated in the form
\begin{equation}
\left[D\vert_{k_{y}=0}+k_{y}\frac{\partial D}{\partial k_{y}}\vert_{k_{y}=0}+\dfrac{1}{2}k_{y}^{2}\frac{\partial^{2}D}{\partial^{2}k_{y}^{2}}\vert_{k_{y}=0}\right]\phi_{1}=0
\end{equation}
to obtain a nonlocal differential equation for the perturbed fields
from the local dispersion equation \ref{dispersionlocal}. On the same token, defining $L_{s}$ as the characteristic magnetic shear scale length and neglecting the ion gravitational Pedersen drift i.e., $V_{0y}=\frac{\nu_{in}}{\Omega_{i}^{2}}\boldsymbol{g}_{\perp y}$
a second order differential equation is obtained as
\begin{multline}
\frac{d^{2}}{dY^{2}}\phi_{1}+\frac{d}{dY}\phi_{1}\\
+\biggl[-\frac{k_{x}^{2}}{\epsilon_{n}^{2}}-\frac{Y^{2}}{L_{s}^{2}\epsilon_{n}^{2}}k_{x}^{2}\frac{\Omega_{i}}{\nu_{in}}\left(\frac{\Omega_{e}}{\nu_{ei}}+\frac{\Omega_{i}}{\nu_{in}}\right)-\frac{\Omega_{i}}{\nu_{in}}\frac{V_{o\perp x}}{\epsilon_{n}^{2}\omega}k_{x}\left(i\epsilon_{n}k_{x}-\frac{\Omega_{e}}{\nu_{ei}}\frac{Y^{2}}{L_{s}^{2}}k_{x}^{2}\right)\\
-\frac{\Omega_{i}}{\nu_{in}}\frac{Y}{L_{s}\epsilon_{n}^{2}}\frac{V_{oz}}{\omega}k_{x}\left(i\epsilon_{n}k_{x}-\frac{\Omega_{e}}{\nu_{ei}}\frac{Y^{2}}{L_{s}^{2}}k_{x}^{2}\right)\biggr]\phi_{1}=0\label{modeequationlong}
\end{multline}
In the derivation of the above equation, the fluctuating
electron Pedersen drift (term proportional to $\left(\frac{\nu_{ei}}{\Omega_{e}}\right)^{2}$
in equation \ref{dispersionlocal}) is neglected. Further ion inertia  term is approximated to $\nu_{in}-i\left(\omega-\boldsymbol{k}.\boldsymbol{V}_{0}\right)\sim\nu_{in}$
and following normalizations have been used
\begin{equation}
Y=y\epsilon_{n};{L_{s}}={l_{s}\epsilon_{n}}
\end{equation}
This equation \ref{modeequationlong} describes the potential eigenmode
structure about a mode rational surface of the general convective
fluid instabilities in an inhomogeneous plasma in the presence of
a parallel current, a transverse zero order electric field and includes
the effect of gravity through $V_{0\perp}$. To solve this for eigenvalues and eigenfunction, the following transformation scheme 
\begin{equation}
\phi_{1}=\psi_{1}exp\left[-\frac{1}{2}\int{dY}\right]
\end{equation}
is used to eliminate the first derivative which further yields Weber like eigenvalue equation for $\psi_{1}$ with complex quadratic potential structure 

\begin{multline}
\frac{d^{2}}{dY^{2}}\psi_{1}+\bigg[-\frac{k_{x}^{2}}{\epsilon_{n}^{2}}-\frac{Y^{2}}{L_{s}^{2}\epsilon_{n}^{2}}k_{x}^{2}\frac{\Omega_{i}}{\nu_{in}}\left(\frac{\Omega_{e}}{\nu_{ei}}+\frac{\Omega_{i}}{\nu_{in}}\right)-\frac{\Omega_{i}}{\nu_{in}}\frac{V_{o\perp x}}{\epsilon_{n}^{2}\omega}k_{x}\left(i\epsilon_{n}k_{x}-\frac{\Omega_{e}}{\nu_{ei}}\frac{Y^{2}}{L_{s}^{2}}k_{x}^{2}\right)\\
-\frac{\Omega_{i}}{\nu_{in}}\frac{Y}{L_{s}\epsilon_{n}^{2}}\frac{V_{oz}}{\omega}k_{x}\left(i\epsilon_{n}k_{x}-\frac{\Omega_{e}}{\nu_{ei}}\frac{Y^{2}}{L_{s}^{2}}k_{x}^{2}\right)+\frac{1}{L_{s}\epsilon_{n}}\frac{\Omega_{i}}{\Omega_{e}}\frac{\nu_{ei}}{\nu_{in}}\frac{V_{oz}}{\omega}k_{x}^{2}-\frac{1}{4}\bigg]\psi_{1}=0\label{modeequationtrandformed}
\end{multline}
Now following \citet{huba1980influence} and substituting $\omega=i\gamma$ in equation \ref{modeequationtrandformed}, one gets

\begin{equation}
A\frac{d^{2}}{dY^{2}}\psi_{1}+\left[Q_{R}+iQ_{I}\right]\psi_{1}=0\label{diffeqcomplex}
\end{equation}
where $A=\epsilon_{n}^{2}/k_{x}^{2}$ and 

\begin{equation}
Q_{R}=-\left[1+\frac{\Omega_{i}}{\nu_{in}}\left(\frac{\Omega_{e}}{\nu_{ei}}+\frac{\Omega_{i}}{\nu_{in}}\right)\frac{Y^{2}}{L_{s}^{2}}+\frac{\epsilon_{n}^{2}}{4k_{x}^{2}}+\frac{V_{o\perp x}}{\gamma}\frac{\Omega_{i}}{\nu_{in}}\epsilon_{n}+\frac{V_{0z}}{\gamma}\frac{Y}{L_{s}}\frac{\Omega_{i}}{\nu_{in}}\epsilon_{n}\right]
\end{equation}

\begin{equation}
Q_{I}=-\left[\frac{V_{o\perp x}}{\gamma}\frac{\Omega_{i}}{\nu_{in}}\frac{\Omega_{e}}{\nu_{ei}}\frac{Y^{2}}{L_{s}^{2}}k_{x}+\frac{V_{oz}}{\gamma}\frac{\Omega_{i}}{\nu_{in}}\frac{\Omega_{e}}{\nu_{ei}}\frac{Y^{3}}{L_{s}^{3}}k_{x}+\frac{V_{oz}}{\gamma}\frac{\nu_{ei}}{\nu_{in}}\frac{\Omega_{i}}{\Omega_{e}}\frac{1}{L_{s}}\epsilon_{n}\right]
\end{equation}
It can be shown that for parameters of ionospheric application $Q_{R}\gg\vert Q_{I}\vert$ \citet{huba1980influence}.
Thus, equation \ref{diffeqcomplex} can be rewritten as,

\begin{equation}
A\frac{d^{2}}{dY^{2}}\psi_{1}+\left[B-C\left(Y-Y_{0}\right)^{2}\right]\psi_{1}=0\label{diffeqmin}
\end{equation}
where

\begin{equation}
B=\frac{\frac{\Omega_{i}}{\nu_{in}}\frac{V_{oz}^{2}}{4\gamma^{2}}\epsilon_{n}^{2}}{\left(\frac{\Omega_{e}}{\nu_{ei}}+\frac{\Omega_{i}}{\nu_{in}}\right)}-\left\lbrace 1+\frac{\epsilon_{n}^{2}}{4k_{x}^{2}}+\frac{V_{o\perp x}}{\gamma}\frac{\Omega_{i}}{\nu_{in}}\epsilon_{n}\right\rbrace 
\end{equation}

\begin{equation}
C=\frac{1}{L_{s}^{2}}\frac{\Omega_{i}}{\nu_{in}}\left(\frac{\Omega_{e}}{\nu_{ei}}+\frac{\Omega_{i}}{\nu_{in}}\right)
\end{equation}

\begin{equation}
Y_{0}=-\frac{1}{2}\frac{V_{oz}}{\gamma}\frac{\epsilon_{n}L_{s}}{\left(\frac{\Omega_{e}}{\nu_{ei}}+\frac{\Omega_{i}}{\nu_{in}}\right)}
\end{equation}
Equation \ref{diffeqmin} is casted in the form of Weber's equation.
Here, $Y_{0}$ is the position of the minimum in the potential well
$Q$. The solution of the above equation \ref{diffeqmin} yields the
\textit{Eigenfunctions} \textit{\emph{in terms of the Hermite polynomials
$H_{l}$}} 
\begin{equation}
\psi_{1}=\psi_{0}H_{l}\left(\left(\frac{C}{A}\right)^{1/4}\left(Y-Y_{0}\right)\right)\exp\left[-\frac{1}{2}\sqrt{\frac{C}{A}}\left(Y-Y_{0}\right)^{2}\right]\label{eigen_f}
\end{equation}
This shows that the eigenmode is localised about the rational surface
due to finite magnetic shear and the mode is shifted off the rational
surface due to equilibrium parallel flow $V_{0z}$ or current $J_{0z}$.
The mode width is simply given by $\Delta=(A/C)^{1/4}$ and mode shift is simply $Y_{0}$. Here, Fig.\,\ref{modecartoonfinal} illustrates a generalised mode-cartoon for the representation of the mode localisation process.

The \emph{eigenfrequency} is obtained from the quantization condition 

\begin{equation}
B=\left(2l+1\right)\left(AC\right)^{\frac{1}{2}}\label{modeeigenfrequency}
\end{equation}
where the radial quantum number $l=\left(0,1,2......\right)$. This
yields the following equation for growth rate of the perturbations
\begin{equation}
\frac{\frac{\Omega_{i}}{\nu_{in}}\frac{V_{oz}^{2}}{4\gamma^{2}}\epsilon_{n}^{2}}{\left(\frac{\Omega_{e}}{\nu_{ei}}+\frac{\Omega_{i}}{\nu_{in}}\right)}-\left\lbrace 1+\frac{\epsilon_{n}^{2}}{4k_{x}^{2}}+\frac{V_{o\perp x}}{\gamma}\frac{\Omega_{i}}{\nu_{in}}\epsilon_{n}\right\rbrace =\left(2l+1\right)\left[\frac{\epsilon_{n}^{2}}{k_{x}^{2}}\frac{1}{L_{s}^{2}}\frac{\Omega_{i}}{\nu_{in}}\left(\frac{\Omega_{e}}{\nu_{ei}}+\frac{\Omega_{i}}{\nu_{in}}\right)\right]^{\frac{1}{2}}
\end{equation}
Simple algebraic manipulations yields the following quadratic equation
for growth rate 

\begin{equation}
P\gamma^{2}+Q\gamma + R=0
\end{equation}
where 

\begin{equation}
P=\left[1+\frac{\epsilon_{n}^{2}}{4k_{x}^{2}}+\left(2l+1\right)\left\lbrace \frac{\epsilon_{n}^{2}}{k_{x}^{2}}\frac{1}{L_{s}^{2}}\frac{\Omega_{i}}{\nu_{in}}\left(\frac{\Omega_{e}}{\nu_{ei}}+\frac{\Omega_{i}}{\nu_{in}}\right)\right\rbrace \right]^{\frac{1}{2}}
\end{equation}

\begin{equation}
Q=\left[V_{o\perp x}\frac{\Omega_{i}}{\nu_{in}}\epsilon_{n}\right]
\end{equation}

\begin{equation}
R=-\frac{\frac{\Omega_{i}}{\nu_{in}}\frac{V_{oz}^{2}}{4}\epsilon_{n}^{2}}{\left(\frac{\Omega_{e}}{\nu_{ei}}+\frac{\Omega_{i}}{\nu_{in}}\right)}
\end{equation}
The global mix-mode growth rate with sheared magnetic field is,  therefore,

\begin{equation}
\gamma_{G}^{S}=\frac{\epsilon_{n}\left(\frac{cE_{0x}}{B_{0}}+\frac{g_{\perp}}{\nu_{in}}\right)+\sqrt{\left\lbrace \epsilon_{n}\left(\frac{cE_{0x}}{B_{0}}+\frac{g_{\perp}}{\nu_{in}}\right)\right\rbrace ^{2}+\frac{\left(\epsilon_{n}\frac{j_{0\parallel}}{n_{0}e}\right)^{2}\left[1+\frac{\epsilon_{n}^{2}}{4k_{x}^{2}}+\frac{\epsilon_{n}}{k_{x}L_{s}}\left\lbrace \frac{\Omega_{i}}{\nu_{in}}\left(\frac{\Omega_{e}}{\nu_{ei}}+\frac{\Omega_{i}}{\nu_{in}}\right)\right\rbrace ^{\frac{1}{2}}\right]}{\left(1+\frac{\Omega_{i}}{\Omega_{e}}\frac{\nu_{ei}}{\nu_{in}}\right)}}}{2\left[1+\frac{\epsilon_{n}^{2}}{4k_{x}^{2}}+\frac{\epsilon_{n}}{k_{x}L_{s}}\left\lbrace \frac{\Omega_{i}}{\nu_{in}}\left(\frac{\Omega_{e}}{\nu_{ei}}+\frac{\Omega_{i}}{\nu_{in}}\right)\right\rbrace ^{\frac{1}{2}}\right]}\label{shear}
\end{equation}
where $l=0$ mode has been considered for simplicity. Further,
in equation \ref{shear}, it is implied that $V_{0z}=\frac{j_{0\parallel}}{n_{0}e}$
and $V_{o\perp x}=\left(\frac{\nu_{in}}{\Omega_{i}}\right) \left(\frac{cE_{0x}}{B_{0}}+\frac{g_{\perp y}}{\nu_{in}}\right)$

\emph{Individual global growth rates in sheared magnetic field:-}

From equation \ref{shear}, expressions could be obtained for the individual growth contributions due to various drivers. First, by putting $V_{0z}=0$ in equation \ref{shear} one could obtain the modified growth rate of the $E \times B$ (and/or collisional Gravitational) instabilities in the presence of a sheared magnetic field. 

\begin{equation}
\gamma_{G-EXB}^{S}=\frac{\epsilon_{n}\left[\frac{cE_{0x}}{B_{0}}+\frac{g_{\perp y}}{\nu_{in}}\right]}{\left[1+\frac{\epsilon_{n}^{2}}{4k_{x}^{2}}+\frac{\epsilon_{n}}{k_{x}L_{s}}\left\lbrace \frac{\Omega_{i}}{\nu_{in}}\left(\frac{\Omega_{e}}{\nu_{ei}}+\frac{\Omega_{i}}{\nu_{in}}\right)\right\rbrace ^{\frac{1}{2}}\right]}
\end{equation}

Similarly,  by putting, $V_{0\perp x}=0$ , one could finally obtain the growth rate for the CCI in the presence of a sheared B-field.
\begin{equation}
\gamma_{G-CCI}^{S}=\frac{\left(\frac{\epsilon_{n}j_{0\parallel}}{n_{0}e}\right)}{2\left(1+\frac{\Omega_{e}}{\Omega_{i}}\frac{\nu_{in}}{\nu_{ei}}\right)^{\frac{1}{2}}\left[1+\frac{\epsilon_{n}^{2}}{4k_{x}^{2}}+\frac{\epsilon_{n}}{k_{x}L_{s}}\left\lbrace \frac{\Omega_{i}}{\nu_{in}}\left(\frac{\Omega_{e}}{\nu_{ei}}+\frac{\Omega_{i}}{\nu_{in}}\right)\right\rbrace ^{\frac{1}{2}}\right]^{\frac{1}{2}}}
\end{equation}

Further the growth rate expression for the nonlocal CCI in the presence of a sheared B-field derived by \citet{huba1980influence}, could be easily recovered.

\emph{Individual local growth rates in shearfree magnetic field:-}

It can be shown that, for instability, by substituting $L_{s}\rightarrow\infty$
in equation \ref{shear}, one could obtain the familiar shear free
local growth rates for the $E\times B$ (and/or Gravitational) instabilities  and CCI \citet{ossakow1979current} by making use of the assumption $\epsilon_{n}\ll k_{x}$. Further, by putting $V_{0z}=0$, one  could obtain the familiar local $E\times B$ and collisional Gravitational instability growth rates obtained in equation \ref{shearfreelocalexb}. Further, by putting $V_{0\perp}=0$, one could regain the maximum growth rate for the CCI as in equation \ref{shearfreelocalcci}.

For the ionospheric application, the parameters given in \citet{huba1980influence} have been used to obtain the growth rate estimations  for the general case of growing perturbations from equation \ref{shear}. The parameters are as follows:- 
$\frac{\Omega_{e}}{\nu_{ei}}\sim\frac{\Omega_{i}}{\nu_{in}}\sim10^{2},\epsilon_{n}^{-1}=L_{N}\sim5\times10^{4}m,
l_{s}\sim3\times10^{6}m,
V_{0\perp x}\sim2\times10^{2}m/sec,
V_{0z}\sim6\times10^{4}m/sec$
Here $l_{s}$ is computed from the Maxwell's equations $\boldsymbol{\nabla}\times\boldsymbol{B}_{0}
=\frac{4\pi}{c}J_{0z}\hat{\boldsymbol{z}}$. Further, we use the classical definition by \citet{kelley1982origin}, which suggests large scale structures to have wavelenghts $\lambda\geq10$ km, intermediate scale in the range $0.1\leq\lambda\leq10$ km, transition wavelengths in the limit $10\leq\lambda\leq100$m, and short wavelengths  $\lambda<10$m.

\begin{figure}[ht!]
\centerline{\includegraphics[scale=0.8]{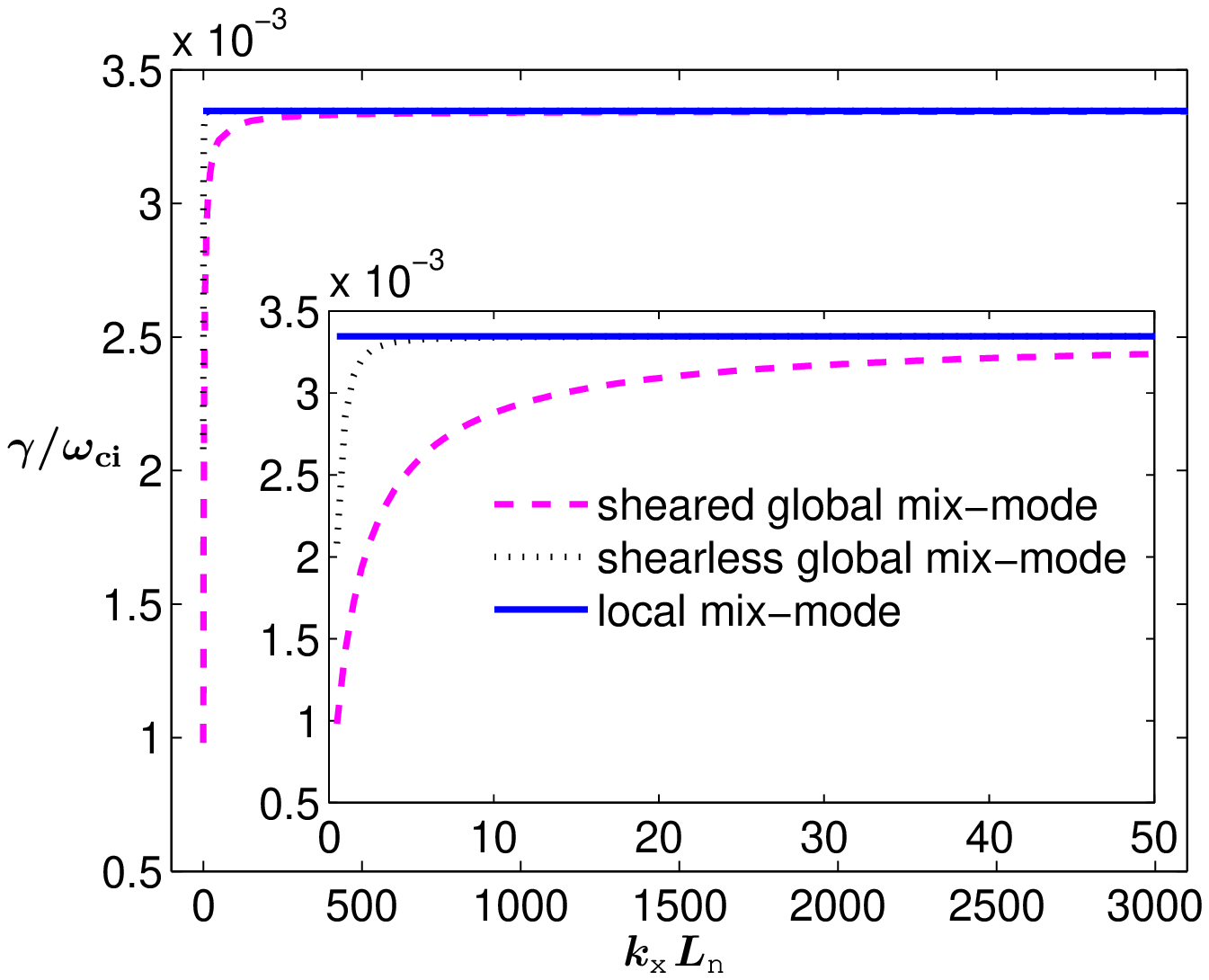}}

\caption{(Colour online) Mix-mode growth rate for the sheared global ( magenta  coloured dashed curves), shearfree global( black coloured dotted curves) and shearfree local cases( blue coloured solid line ). The zoomed image is given in inset.}
\label{mixfinal}
\end{figure}

\begin{figure}[ht!]
\centerline{\includegraphics[scale=0.8]{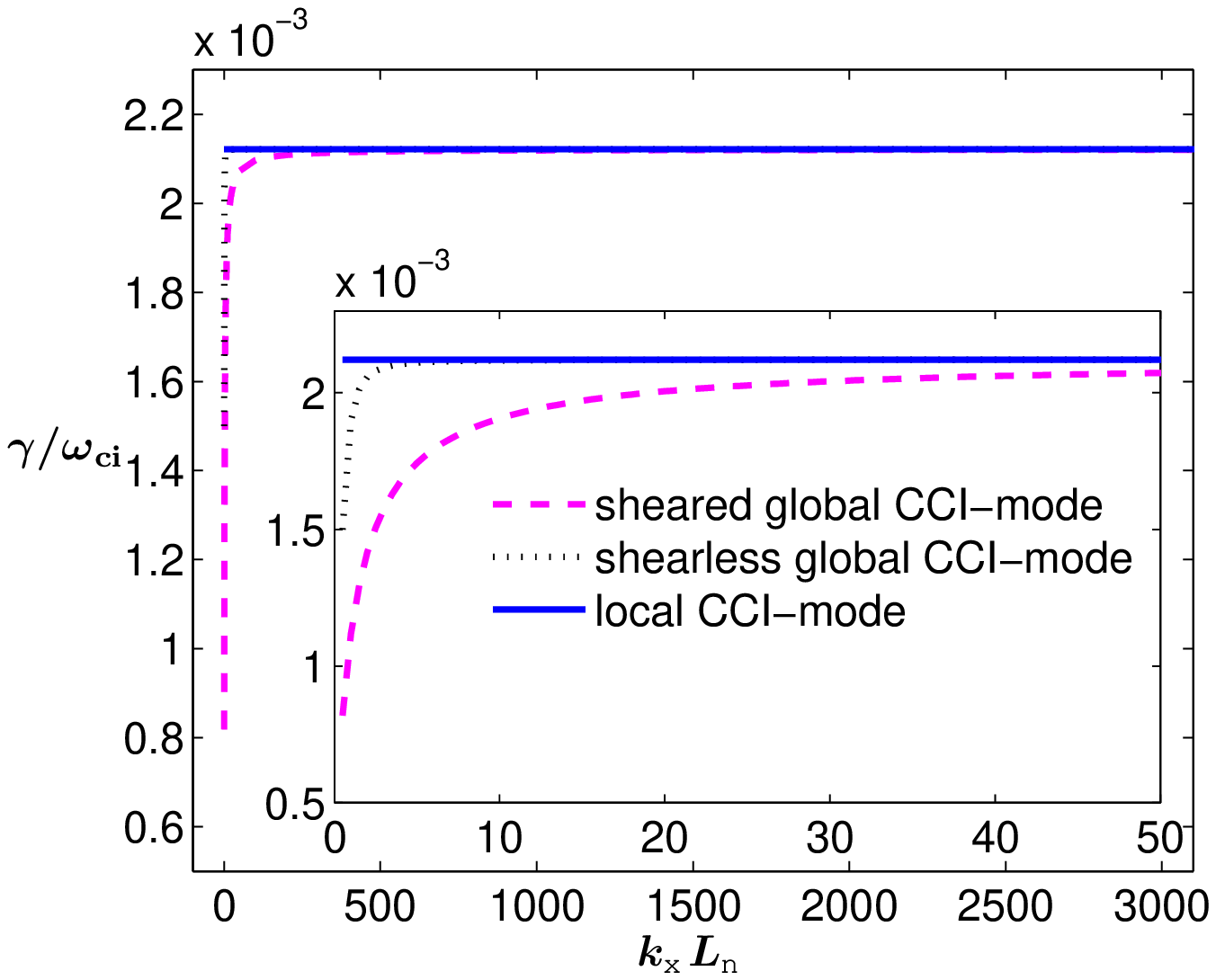}}

\caption{(Colour online) CCI growth rate for the sheared global (magenta  coloured dashed curves), shearfree global (black coloured dotted curves) and shearfree local cases (blue coloured solid line). The zoomed image is given in inset}
\label{ccifinal}
\end{figure}

\begin{figure}[ht!]
\centerline{\includegraphics[scale=0.8]{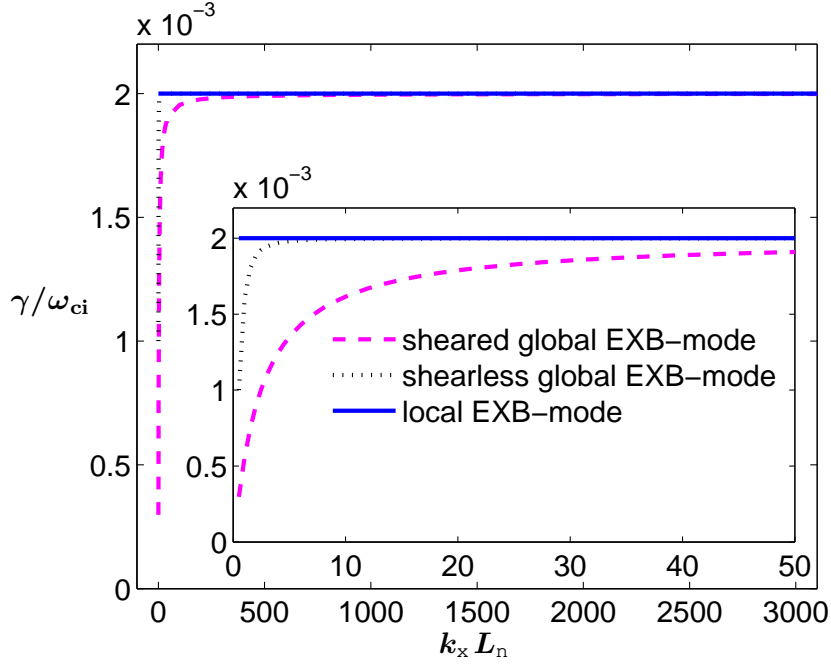}}

\caption{(Colour online) $E\times B$ growth rate for the sheared global (magenta coloured dashed curves), shearfree global (black  coloured dotted curves) and shearfree local cases (blue coloured solid line). The zoomed image is given in inset}
\label{exbfinal}
\end{figure} 

Fig.\,\ref{mixfinal},\,\ref{ccifinal} and \,\ref{exbfinal} represents the indiviual growth rate estimations of the sheared global, shearfree global and shearfree local cases for mix-mode, CCI and $E\times B$ (and/or Gravitational) instabilities respectively. Here, the growth rates of the three types of instabilities have been plotted as a function of wavelengths. It is clear that at larger scale sizes, the global sheared growth rate shows significant reduction from the corresponding shearless local growth rate values. Another important aspect is the reduction of global sheared growth rates in comparision to shearless global growth rates at larger scale sizes.

However, at the intermediate scale sizes ( $\sim$ a few kms), that are interesting from the practical point of view since they are responsible for the scintillation of satellite signals, the sheared global growth
rates are reduced only by a small factor than the corresponding shearless local growth rate values. In other words, at scintillation causing scale sizes the growth is lagely unaffected for these instabilities, which is in agreement with the work of \citet{huba1980influence} exclusively done for the CCI in the influence of magnetic shear.

\section{Conclusion}
Recently, \citet{burston2016polar} investigated the role of four primary plasma instabilties associated with the polar cap plasma patch. In that work, GDI, CCI, KHI and small scale turbulence processes were examined statistically to account the generation of phase scintillation associated with the polar cap plasma patches using Dynamic Explorer 2 Satellite data. On the same lines of thought, the effect of magnetic field shear is investigated analytically for the $E\times B$ (and/or Gravitational) and the CCI to analyse local and global fluid flow patterns in slab geometry configuration. Further, the global mix-mode potential eigen mode yield the mode localisation and shift signatures. It turns out that the mode is localised about the rational surface due to finite magnetic shear while it gets shifted off the rational surface due to equilibrium parallel flow or ambient current in the system. It turns out that the magnetic shear induced stabilization is more effective at the larger scale sizes while at the intermediate scintillation causing scale sizes, the growth is largely unaffected for these instabilities. 

Further, these results supplement and extend the earlier work done exclusively for the CCI under the influence of magnetic shear to account for a more realistic polar ionospheric plasma patch morphology \citet{huba1980influence}. Moreover, various parameters such as $\nu_{ei},\nu_{in},V_{0z},V_{0\perp x}$ , are all variables in the actual ionosphere with respect to the altitude (Z-axis), latitude (Y-axis) and longitude (X-axis) and adversely intensify/weaken the localised  scintillation processes. Simulation studies need to be include these effects for a more realistic scenario. Finally, convective instabilities such as CCI and $E\times B$ (and/or Gravitational) instabilities are likely to play an important role in structuring the medium through the critical analysis of total electron content. Thus, it turns out that these studies will provide more information for local and global variation of electron density profiles as well as other auxilliary nonlinear plasma parameters which are needed to improve the accuracy and precision of the 4-D ionospheric imaging algorithms and scintillation forecast models \citet{bust2008history}, \citet{burston2012imaging}.

\acknowledgments 
One of the authors (Sushil Kumar Singh) acknowledges
Pradeep Kumar Chaturvedi( formerly at Naval Research Laboratory and Maryland University,U.S.A) for useful discussions at Centre for Energy Studies,I.I.T.Delhi. Author (Jyoti k Atul) acknowledges
Vikrant Saxena for useful discussions at I.P.R. Gandhinagar.

\bibliographystyle{agu}


\end{document}